\documentclass[12pt]{iopart}
\usepackage{graphicx}
\usepackage{natbib} 
\usepackage[dvipsnames]{xcolor}
\usepackage{soul}
\usepackage{hyperref}

\expandafter\let\csname equation*\endcsname\relax
\expandafter\let\csname endequation*\endcsname\relax

\usepackage{amsmath}
\usepackage{amssymb}
\usepackage{txfonts}

\usepackage[utf8]{inputenc}
\usepackage[T1]{fontenc}

\usepackage[polish,english]{babel}
\usepackage[]{algorithm2e}
\newcommand{\F}{{\mathcal F}}

\newcommand{\be}{\begin{equation}}
\newcommand{\ee}{\end{equation}}
\newcommand{\bea}{\begin{eqnarray}}
\newcommand{\eea}{\end{eqnarray}}

\newcommand{\tav}[1]{\left\langle#1\right\rangle}

\newcommand{\bdm}{\begin{displaymath}}

\newcommand{\bse}{\begin{subequations}}

\newcommand{\btheta}{\boldsymbol{\theta}}
\newcommand{\bxi}{\boldsymbol{\xi}}

\newcommand{\mi}{\mathrm{i}}

\newcommand{\sfA}{\mathsf{A}}

\newcommand{\sfF}{\mathsf{F}}

\newcommand{\sfM}{\mathsf{M}}

\newcommand{\sfS}{\mathsf{S}}

\newcommand{\sfT}{\mathsf{T}}

\newcommand{\sfa}{\mathsf{a}}
\newcommand{\sfh}{\mathsf{h}}

\newcommand{\edm}{\end{displaymath}}

\newcommand{\ese}{\end{subequations}}

\newcommand{\mFr}{\widetilde{\Gamma}}

\newcommand{\calA}{\mathcal{A}}
\newcommand{\calB}{\mathcal{B}}

\newcommand{\pa}{\partial}

\begin{document}

\title[Follow-up procedure for the time-domain $\F$-statistic method]{Follow-up procedure for gravitational wave searches from isolated neutron stars using the time-domain $\F$-statistic method} 
\author{Magdalena Sieniawska$^1$, Micha\l{} Bejger$^{1,2}$, Andrzej Kr\'{o}lak$^3$}

\address{$^1$ Nicolaus Copernicus Astronomical Center, Polish Academy of Sciences, Bartycka 18, 00--716 Warszawa, Poland}
\address{$^2$ APC, AstroParticule et Cosmologie, Université Paris Diderot, CNRS/IN2P3, CEA/Irfu, Observatoire de Paris, Sorbonne Paris Cité, F-75205 Paris Cedex 13, France} 
\address{$^3$ Institute of Mathematics, Polish Academy of Sciences, \'Sniadeckich 8, 00--656 Warszawa, Poland}

\ead{msieniawska@camk.edu.pl}


\begin{abstract}

Among promising sources of gravitational waves are long-lived nearly periodic signals produced by rotating, asymmetric neutron stars. Depending on the astrophysical scenario, the sources of asymmetry may have thermal, viscous, elastic and/or magnetic origin. In this work we introduce a follow-up procedure for an all-sky search for gravitational wave signals from rotating neutron stars. The procedure denoted as \texttt{Followup} implements matched-filtering {$\F$}-statistic method. We describe data analysis methods and algorithms used in the procedure. We present tests of the
\texttt{Followup} for artificial signals added to white,
Gaussian noise. The tests show a good agreement with the theoretical predictions. The \texttt{Followup} will become
part of the \texttt{Time-Domain $\F$-statistic} pipeline that is routinely used for all-sky searches of LIGO and Virgo detector data.

\end{abstract}

\submitto{CQG}
\maketitle

\section{Introduction}
\label{sect:intro} 

Historically the first gravitational-wave (GW) signal 
registered on Earth \citep{Abbott2016a} was caused by a merger of a stellar-mass black hole binary system. Since then nine more such events were observed \citep{Abbott2016b, Abbott2017a, Abbott2017b, Abbott2017c, Abbott2018b}
by LIGO and Virgo laser interferometric detectors. Additionally, one binary neutron star (NS) merger \citep{Abbott2017d} was detected.

Due to the unremitting efforts of the LIGO and Virgo \citep{Acernese2014} Collaboration (LVC) in the detectors upgrades and in improvements of the data analysis methods, it will be possible to detect much subtler signals, in particular those emitted by rotating asymmetric neutron stars. 
Several searches for continuous gravitational waves (CGW)   from isolated neutron stars have been carried out in LIGO and Virgo data (see \citealt{Riles2017} for a recent review). These searches have included coherent searches for gravitational radiation from known radio and X-ray pulsars, directed searches for known stars or locations having unknown signal frequencies, and spotlight or all-sky searches for stars with unknown signal frequency and sky location.
Even though no statistically-significant signals were detected, interesting upper limits on the CGW emission were placed. CGW are expected to have strain amplitudes a few orders of magnitude weaker than signals produce by mergers. However in contrast to the merger signals, CGW are long-lasting, and therefore one can improve the sensitivity of the searches by increasing the observational time.

According to the current theoretical state-of-art, several mechanisms may be responsible for a long-lasting GW. In the case of young and isolated neutrons stars (NS), strong, evolving toroidal magnetic field and unstable oscillation modes (r-modes, \citealt{Owen1998}) may lead to detectable signals. NS in accreting binaries may accumulate accreted material on their surface. Deformation due to mass, temperature gradients, elastic or magnetic field strain result in a non-axisymmetric shape of NS and CGW emission \citep{Bonazzola1996, Bildsten1998, Ushomirsky2000}. In general, any rotating non-axisymmetric NS (exhibiting time-varying mass quadrupole moment) will produce GW; for the review see \citet{Andersson2011,Lasky2015,Riles2017}. Canonical emitters of CGW and main targets for the LVC searches are rapidly-rotating, non-axisymmetric NS in our Galaxy. Their model is motivated by the relation between GW strain amplitude $h_0$ and the spin frequency $f$ \citep{Zimmerman1979}:
\begin{equation}
h_0=4 \times 10^{-25} \left( \frac{\epsilon}{10^{-6}} \right) \left( \frac{I_3}{10^{45} \textrm{ g cm}^2} \right) \left( \frac{f}{100\textrm{ Hz}} \right)^2 \left( \frac{100 \textrm{ pc}}{d} \right),
\end{equation}
where $d$ is the distance to the source, $\epsilon = (I_1 - I_2)/I_3$ and $I_1$, $I_2$, $I_3$ denote moments of inertia along three axes, with the direction of $I_3$ aligned with the axis of the NS angular momentum.

One of the pipelines to perform all-sky search for gravitational waves from isolated rotating neutron stars 
is \texttt{Time-Domain $\F$-statistic} pipeline\footnote{Project's repository: \url{https://github.com/mbejger/polgraw-allsky}}. 
This pipeline implements a matched-filter statistic for detecting nearly periodic signals in the time-domain data, called the $\F$-statistic \citep{Jaranowski1998, Jaranowski2009, Jaranowski2010, Astone2010, Pisarski2015}. 
The \texttt{Time-Domain $\F$-statistic} pipeline is divided into several steps: generation of time domain data, construction of grid of templates, $\F$-statistic search for candidate signals, search for coincidences between candidates in different time segments, calculation of false alarm probability for coincidences obtained, estimation of sensitivity of the search if no significant signal is detected. 

There are several other all-sky search pipelines that are used in the searches for gravitational waves from isolated neutron stars in LIGO and Virgo detectors data: \texttt{Einstein@Home} \citep{Abbott2009} \footnote{\url{http://einstein.phys.uwm.edu}}, \texttt{FrequencyHough} \citep{Astone2014}, \texttt{PowerFlux} \citep{Abbott2008}, and \texttt{SkyHough} \citep{Krishnan2004} (for a recent review see \citealt{Bejger2017}). 

In this paper we present a follow-up procedure which aims at
verifying whether promising candidates obtained by 
the \texttt{Time-Domain $\F$-statistic} pipeline can be of astrophysical origin. The follow-up procedure is based on the assumption that gravitational wave signal is a coherent signal that is always present in the data. Consequently when coherence time for matched-filtering analysis increases the signal-to-noise ratio increases as square root of the observational time.


The article is composed as follows: in Sect.~\ref{sect:polgraw} we describe the model of the expected CGW signal and the $\F$-statistic method. In Sect.~\ref{sect:method} we introduce the \texttt{Followup} procedure: description of steps (Sect.~\ref{sect:steps}), construction of the optimal grid of templates (Sect.~\ref{sect:OptGrid}) and optimisation algorithms used to find maximum of the $\F$-statistic over the parameter space (Sect.~\ref{sect:maximum}). In Sect.~\ref{sect:testsGN} we present implementation of the \texttt{Followup} procedure for the of CGW signal buried in white Gaussian noise. We perform Monte Carlo simulations for two cases: the two-dimensional parameter space (Sect.~\ref{sect:test2d}), as well for general, four-dimensional parameter space (Sect.~\ref{sect:test4d}). Sect.~\ref{sect:summ}, summarizes and discusses application of the \texttt{Followup} procedure to the \texttt{Time-Domain $\F$-statistic} pipeline.

\section{CGW signal model and the Time-Domain $\F$-statistic method}
\label{sect:polgraw}

The \texttt{Followup} algorithm is a part of the \texttt{Time-Domain $\F$-statistic} all-sky search pipeline. The pipeline performs a coherent search for CGW signals in the time-domain data segments using the $\F$-statistic \citep{Jaranowski1998}. Here, for the context, we provide a brief introduction to this method.

The time-domain response $s(t)$ of the interferometric detector to a CGW signal is given by a linear combination of four time-dependent components:
\be
\label{sig}
s(t) = \sum^4_{i=1} a_{i}\,h_{i}(t),
\ee
where the functions $h_i$ ($i=1,\dots,4$) are of the form
\be
\label{eq:ab}
\begin{array}{c}
h_1(t) = a(t)\cos\phi(t),\quad h_2(t) = b(t)\cos\phi(t),
\\[2ex]
h_3(t) = a(t)\sin\phi(t),\quad h_4(t) = b(t)\sin\phi(t),
\end{array}
\ee
The functions $a(t)$ and $b(t)$ are the amplitude modulation functions that depend on the location and orientation 
of the detector on the Earth and on the position of the GW source in the sky, 
described in the equatorial coordinate system by the right ascension $\alpha$ and the declination $\delta$ angles. 
They are periodic functions of time with the period of one and two sidereal days. The analytic form of the functions 
$a(t)$ and $b(t)$ for the case of interferometric detectors is given by Eqs.\ 12, 13 of \citet{Jaranowski1998}.
The phase $\phi(t)$ is given by
\begin{equation}
\label{eq:phase}
\phi(t) = \omega t + \dot{\omega} t^2 + \frac{{\bf n} \cdot {\bf r}_d(t)}{c} (\omega + 2 \dot{\omega} t),
\end{equation}
where ${\bf r}_d(t)$ is the vector that joins the solar-system barycenter (SSB) with the detector, and ${\bf n}$ is the unit vector pointing from the SSB to the source. In equatorial coordinates
$(\delta$, $\alpha)$ we have ${\bf n} = (\cos\delta\cos\alpha,\cos\delta\sin\alpha,\sin\delta)$. In the following we assume that the frequency evolution is accurately described by one spindown parameter $\dot{\omega}$.
The four amplitudes $a_{i}$ depend on amplitude $h_0$, phase
$\phi_0$, polarization angle $\psi$, and inclination angle
$\iota$ (see Eqs.\ 2.10 of \citealt{Astone2010}).

CGW signals are identified with the use of the $\F$-statistic introduced in \citet{Jaranowski1998}. The $\F$-statistic is obtained by maximizing the likelihood function with respect to the four unknown parameters - $h_0$, $\phi_0$, $\iota$, and $\psi$ (which are henceforth called the \textit{extrinsic} parameters). This leaves a function of only - in the most basic case - four remaining parameters: $\omega$, $\dot{\omega}$, $\delta$, and $ \alpha$ (henceforth called the \textit{intrinsic} parameters). The angular frequency $\omega$ and angular frequency derivative $\dot{\omega}$ are dimensionless and they are defined by $\omega = 2\pi f dt$ and $\dot{\omega} = 2 \pi \dot{f} dt^2$, where $dt$ is sampling time. Thus the dimension of the parameter space that we need to search decreases from 8 to 4.
Additionally, we introduce several simplifications. We set the observation time $T_0$ equal to the integer multiple of sidereal days. We also assume that 
the noise in the detector is white. This is a good approximation because we can assume that over a very narrow band of the signal spectral density of the noise of the detector is constant. 
We assume that data $x(t)$ is a discrete time series
consisting of $N$ uniformly sampled data points,
i.e., $t = 1, ... , N$.
We also introduce a product $\tav{\cdot}$ defined as
\be 
\tav{g(t)} = \frac{1}{\sigma^2}\sum^N_{t=1} g(t),
\ee 
where $\sigma^2$ is the variance of the noise.

Under these assumptions the  $\F$-statistic is given by
\be
\label{eq:Fstat}
\F =
\frac{|F_a|^2}{\tav{a^2(t)}} + \frac{|F_b|^2}{\tav{b^2(t)}},
\ee
where
\bea
\label{Fab}
F_{a} := \tav{x(t)\, a(t) \exp[-\mi\phi(t)]},
\\ \nonumber
F_{b} := \tav{x(t)\, b(t) \exp[-\mi\phi(t)]}.
\eea

The signal-to-noise ratio $\rho$ is defined as
\be
\label{snr}
\rho^2 = \tav{s^2(t)}.
\ee
For Gaussian noise $\rho$ determines probability of the detection 
of signal $s(t)$.
Assuming Gaussian noise the Fisher matrix for signal $s(t)$ 
is given by
\be
\label{mFisher}
\Gamma_{{\theta_i}{\theta_j}} = 
\tav{\frac{\partial s(t)}{\partial\theta_i}\frac{\partial s(t)}{\partial\theta_j}},
\quad i,j=1,\ldots,M.
\ee
where $\theta = (\theta_1,\ldots,\theta_M)$ are $M$ parameters of the signal $s(t)$. 
For sufficiently high signal-to-noise ratio the accuracy of the determination of parameters is approximately 
given by the covariance matrix which is equal to the inverse of the Fisher matrix. As we shall see in the following the Fisher matrix is also useful in the construction of the grid
of templates.
Often it is convenient to consider twice the value of the
$\F$-statistic. This is because $2 \times \F$ has a central $\chi^2$ distribution 
with 4 degrees of freedom when the signal is absent and non-central $\chi^2$ distribution 
with 4 degrees of freedom and non-centrality parameter $\rho^2$ when the signal is present (see Section IIIB of \citealt{Jaranowski1998}). 
Thus when the signal is present  mean ($\mu$) and variance ($\Sigma^2$) of $2\times\F$ read
\bea
\label{eq:mu}
\mu &=& n + \rho^2, \\
\Sigma^2 &=& 2(n + 2\rho^2),
\eea
with $n = 4$.


A fully-coherent search over the whole set of data (typically of the length of a year) is computationally prohibitive. In order to alleviate this problem we apply a semi-coherent method, 
which consists of dividing the data into shorter time domain segments. The short time domain data are analysed coherently 
with the $\F$-statistic. Moreover, to reduce the computer memory required to do the search, the data are divided 
into narrow-band segments that are analysed separately. For a typical search we choose the length of the time-domain segment 
to be several sidereal days long and the bandwidth of the narrow band segment to be a fraction of $1$ Hz. Consequently 
the Time-Domain $\F$-statistic all-sky search pipeline consists of two parts. The first part is a coherent search of narrow-band time domain data segments, where we search a 4-parameter space defined by angular frequency
$\omega$, angular frequency derivative $\dot{\omega}$, declination $\delta$, and right ascension
$\alpha$. The search is performed on an \textit{effective} 4-dimensional grid in the parameter space described in detail in \citet{Astone2010,Pisarski2015}. We set a fixed threshold for
the  $\F$-statistic for each data segment. All the threshold crossings are recorded together with corresponding 4 parameters of the grid point and the signal-to-noise ratio $\rho_c$ at the threshold crossing. The signal-to-noise ratio $\rho_c$ that we record is defined in terms of the $\F$-statistic value $\F_c$ at the threshold crossing as
\be
\label{eq:F2snr}
\rho_c = \sqrt{2 (\F_c - 2)}.
\ee
The formula follows from Eq.~\ref{eq:mu} where for mean
value $\mu$ we substitute $2 \times \F_c$. 

In this way for each time domain segment we obtain a set of candidates.  The second part of the analysis consists of the search for coincidences among the candidates from different time segments. The coincidence procedure is described in detail in Section 8 of \citet{Aasi2014}. We estimate the statistical significance of the coincidences by calculating the probability
that a given coincidence is by chance only. The formula for this false alarm probability is given in the Appendix of \citet{Aasi2014}. Whenever
false alarm probability is sufficiently small we mark the coincidence as significant. Typically we choose the false alarm probability
equal to $10^{-3}$. When we have a candidate coincident 
in $l$ time domain segments we estimate the parameters of 
the coincident candidate as the mean of the parameters of the
$l$ candidates from individual time frames entering the coincidence. The significant candidates are then  subject to the follow-up procedure which is presented in this paper.

So far, the pipeline has been used in several all-sky searches for CGW signals: Virgo VSR1 data \citep{Aasi2014}, LIGO O1 data \citep{Abbott2017e, Abbott2018a}, LIGO O2 data \citep{Abbott2019} and in the LIGO S6 Mock Data Challenge \citep{Walsh2016}.

In the case of an all-sky search, when large parameter space $(\omega, \dot{\omega}, \delta, \alpha)$ has to be investigated scrupulously, optimal usage of the computational resources and reduction of the computational costs play an important role. On the one hand computational cost of the all-sky search depends on the observation time $T_0$ as ${\propto}$\,$T_0^5\log T_0$ \citep{Astone2010}. On the other hand the signal-to-noise ratio scales as $\rho\,{\propto}\sqrt{T_0}$. These two scaling relations show how important balance between sensitivity and computational power is: in the too short data segments signal will be buried deeply in the noise and impossible to restore, while too long time series require unreachable computational power. Interplay between reduction of numerical cost and sensitivity loss indicate that search settings, like e.g. length of the data segments, sampling time, density of the search grid etc., have to be chosen carefully.

\section{The \texttt{Followup} procedure}
\label{sect:method} 

The aim of the \texttt{Followup} procedure is to verify whether significant coincident candidates are of astrophysical interest. 
We assume that the signal's lifetime is longer than the duration of the observing run and consequently we assume that
a signal of astrophysical origin will always be present in the data analysed. From the two-step search procedure of the
\texttt{Time-Domain $\F$-statistic} all-sky pipeline 
we have candidate signals present in  \textit{L} time domain segments, each of \textit{D} sidereal days long. As a coherent signal of true astrophysical origin should be present in all the data i.e. in \textit{L} $\times$ \textit{D} sidereal days long, the signal-to-noise ratio should increase by factor $\sqrt{L}$ with respect to the signal-to-noise in one segment. In addition the accuracy of the estimation of parameters of the signal should also increase considerably. The exact increase in accuracy is complicated by the fact that parameters of the signal are correlated. To facilitate the task of following up the signal we implement a hierarchical procedure. Namely we first search for a coherent signal (evaluate the $\F$-statistic) in two concatenated time-domain segments, then in four segments and so on.

\subsection{Steps of the \texttt{Followup} procedure}
\label{sect:steps}

The candidate signal summarized in Sect.~\ref{sect:polgraw} is described by
four parameters: $\omega_i$, $\dot{\omega}_i$, $\delta_i$, $\alpha_i$.

The steps of the \texttt{Followup} procedure are the following: 
\begin{enumerate}
\item
Take two adjacent time domain segments of the band where the candidate signal is present. 
\item
For each segment construct an \textit{optimal grid} (see Sect.~\ref{sect:OptGrid}) around the parameters of the candidate signal.
\item
Perform a {\it coarse} search by evaluating the $\F$-statistic on the grid, find the maximum value of $\F$, and the corresponding coarse estimates of signal parameters:
$\omega_c$, $\dot{\omega}_c$, $\delta_c$, $\alpha_c$. 
\item
Using an \textit{optimization algorithm} (see Sect.~\ref{sect:maximum}) perform a {\it fine} search for the maximum of the $\F$-statistic and find the corresponding refined estimates of signal parameters. Take an average of the refined signal parameters from the two segments to obtain the {\it fine} estimates $\omega_f$, $\dot{\omega}_f$, $\delta_f$, $\alpha_f$ of four signal parameters. 
\item
Join the two segments together to obtain one segment of double length, and construct the \textit{optimal grid} around the fine values of the parameters
$\omega_f$, $\dot{\omega}_f$, $\delta_f$, $\alpha_f$ obtained in the previous step. Perform again a coarse and a fine search. This will result
in the final \texttt{Followup} parameters $\omega_{fin}$, $\dot{\omega}_{fin}$, $\delta_{fin}$, $\alpha_{fin}$. 
\end{enumerate}

At the end of the procedure, the SNR of a signal of astrophysical origin should increase approximately by a factor of $\sqrt{2}$ with respect to its SNR in individual segments. 
Also the estimates $\omega_{fin}$,  $\dot{\omega}_{fin}$, $\delta_{fin}$, $\alpha_{fin}$ should be more accurate estimates of the parameters of the signal than the initial estimates $\omega_i$, $\dot{\omega}_i$, $\delta_i$, 
$\alpha_i$.
We can then iterate the above procedure to further join the segments. For a true signal the SNR should continue to increase with the above-mentioned factor.

\subsection{Optimal grid}
\label{sect:OptGrid}

The grid used in the coherent $\F$-statistic search was introduced in \citet{Astone2010} and was further optimized in \citet{Pisarski2015}. This grid is optimal for an approximate linear model of the CGW signal
(see Sections IIIB and IV of \citealt{Astone2010}). 
The resulting grid is uniform, which simplifies the search procedure considerably. Moreover, the grid is constrained so that the $\omega$ grid points coincide with the Fourier frequencies. This enables the use of the FFT algorithm and results in computational speed-up. 

In the \texttt{Followup} procedure we are less computationally-bound and we can use a fully optimal grid based on the \textit{reduced} Fisher matrix introduced in \citet{Jaranowski2005}. 
Let us summarize the steps of the construction of the 
\textit{reduced} Fisher matrix (for more details see 
Section 4.4.1 of \citealt{Jaranowski2005} and  Chapter 6 of \citealt{Jaranowski2009}). We present the 
construction for the general case of $n$ amplitude parameters
and $m$ intrinsic parameters.
First we introduce the following shorthand notation.
Let us collect the $n$ amplitude parameters $a_{i}$
and the $n$ waveforms $h_i$ into column vectors.
\be
\label{eq:ah}
\sfa := \begin{pmatrix}a_{1} \\
\vdots \\ a_{n} \end{pmatrix},
\quad
\sfh(t;\bxi) := \begin{pmatrix}h_1(t;\bxi) \\
\vdots \\ h_n(t;\bxi) \end{pmatrix}.
\ee
With this notation the signal $s(t)$
can compactly be written in the following form:
\be
\label{eq:gsig}
s(t;\btheta) = \sfa^\sfT \cdot \sfh(t;\bxi),
\ee
where $\btheta$ is the set of signal parameters, $\sfT$ stands for the matrix transposition and $\cdot$ denotes matrix multiplication. The the $n + m$ signal parameters $\btheta$
consist of $n$ extrinsic amplitude parameters $\sfa$  and $m$ intrinsic parameters $\bxi$. 
For the case of the gravitational wave signal from a rotating neutron star presented in Section \ref{sect:polgraw} there are
$n = 4$ amplitude parameters and 4 intrinsic parameters,
$\bxi = (\omega, \dot{\omega}, \delta, \alpha)$. 

The Fisher matrix $\Gamma$ for the general signal $s(t;\btheta)$ can be written in terms of block matrices as 
\be
\label{G}
\Gamma(\sfa,\bxi) = \begin{pmatrix}
\Gamma_{\sfa\sfa}(\bxi)
& \Gamma_{\sfa\bxi}(\sfa,\bxi)
\\[1ex]
\Gamma_{\sfa\bxi}(\sfa,\bxi)^\sfT
& \Gamma_{\bxi\bxi}(\sfa,\bxi)
\end{pmatrix},
\ee
where $\Gamma_{\sfa\sfa}$ is an $n\times n$ matrix with components
$\tav{{\pa s}/{\pa a_i}\,{\pa s}/{\pa a_j}}$ $(i,j=1,\ldots,n)$,
$\Gamma_{\sfa\bxi}$ is an $n\times m$ matrix with components
$\tav{{\pa s}/{\pa a_i}\,{\pa s}/{\pa\xi_\calA}}$ $(i=1,\ldots,n,\,\calA=1,\ldots,m)$,
and $\Gamma_{\bxi\bxi}$ is an $m\times m$ matrix with components
$\tav{{\pa s}/{\pa\xi_\calA}\,{\pa s}/{\pa\xi_\calB}}$ $(\calA,\calB=1,\ldots,m)$.
The explicit form of these matrices is 
\bse
\label{Gbb}
\begin{align}
\Gamma_{\sfa\sfa}(\bxi) &= \sfM(\bxi),
\\[2ex]
\Gamma_{\sfa\bxi}(\sfa,\bxi) &= \begin{pmatrix}
\sfF_{(1)}(\bxi) \cdot \sfa & \cdots & \sfF_{(m)}(\bxi) \cdot \sfa
\end{pmatrix},
\\[2ex]
\Gamma_{\bxi\bxi}(\sfa,\bxi) &= \begin{pmatrix}
\sfa^\sfT \cdot \sfS_{(11)}(\bxi) \cdot \sfa & \cdots
& \sfa^\sfT \cdot \sfS_{(1m)}(\bxi) \cdot \sfa \\
\hdotsfor{3} \\[1ex]
\sfa^\sfT \cdot \sfS_{(m1)}(\bxi) \cdot \sfa & \cdots
& \sfa^\sfT \cdot \sfS_{(mm)}(\bxi) \cdot \sfa
\end{pmatrix}.
\end{align}
\ese
The components of the matrix $M$ are given by
\be
\label{sec7.2c}
M_{ij}(\bxi) := \tav{h_i(t;\bxi)h_j(t;\bxi)},  \quad  i,j = 1,\ldots,n.
\ee
The components of $m$ matrices $\sfF_{(\calA)}$ 
(note here the index $\calA$ within parentheses has a meaning of the matrix label), and the components of $m^2$ matrices $\sfS_{(\mathcal{AB})}$ read
\bse
\label{sec7.2d}
\be
\label{sec7.2d1}
F_{(\calA){ij}}(\bxi)
:= \bigg< h_i(t;\bxi) \,
\frac{\pa h_j(t;\bxi)}{\pa\xi_\calA} \bigg>,
\quad \calA=1,\ldots,m, \quad i,j=1,\ldots,n,
\ee

\be
\label{sec7.2d2}
S_{(\calA\calB){ij}}(\bxi)
:= \bigg< \frac{\pa h_i(t;\bxi)}{\pa\xi_\calA} \,
\frac{\pa h_j(t;\bxi)}{\pa\xi_\calB} \bigg>,
\quad \calA,\calB=1,\ldots,m, \quad i,j=1,\ldots,n, 
\ee
\ese
respectively. We then define a $m\times m$ square matrix $\mFr$ with components
\be
\label{eq:rF}
\mFr_{\calA\calB}(\bxi) := \frac{1}{n} \text{Tr} \Big(
\sfM(\bxi)^{-1} \cdot \sfA_{(\calA\calB)}(\bxi) \Big), \quad
\calA,\calB=1,\ldots,m,
\ee
where $\sfA_{(\calA\calB)}$ is an $n\times n$ matrix defined as
\be
\label{sec7.2g}
\sfA_{(\calA\calB)}(\bxi) := S_{(\calA\calB)}(\bxi)
- \sfF_{(\calA)}(\bxi)^\sfT \cdot \sfM(\bxi)^{-1} \cdot \sfF_{(\calB)}(\bxi) ,
\quad \calA,\calB=1,\ldots,m.
\ee
We shall call $\mFr$ the \textit{reduced Fisher matrix}.
This matrix is a function of the intrinsic parameters alone and is used to construct a grid on the intrinsic parameter space. 

The construction of a grid in the parameter space is equivalent to a mathematical problem known as the {\em covering problem} \citep{Con}.
In the Euclidean space the covering problem is to cover the space by hyperspheres of radius $R$ in such a way that any point of the space belongs to at least one hypersphere. The radius $R$ is called the {\em covering radius}.
The covering thickness $\Theta$ is defined as the average number of hyperspheres that contain a point in the space. The {\em optimal covering} has the minimal possible thickness. Thus the optimal covering is also referred to as the thinnest
(see Section IV of \citealt{Astone2010} for more details). 

Let $M_o$ be the generator matrix of the thinnest covering in the Euclidean space. To obtain the generator matrix $M_g$ in the intrinsic parameter space we need to transform $M_g$ to space spanned by intrinsic parameters $\bxi$.
This is obtained in the following way
\be
M_g = M_o \cdot V' \cdot \sqrt{E},
\ee
where $V$ is the matrix whose columns are eigenvectors of the reduced Fisher matrix $\mFr$, $V'$ denotes transpose of matrix $V$ and $E$ is the diagonal matrix whose diagonal components are eigenvalues of $\mFr$
(see Section IV of \citealt{Astone2010} for more details).

The constructed grid depends on the values of the intrinsic parameters, i.e. it changes throughout the parameter space. However, in a small neighbourhood around the parameters of the candidate signal the changes are small, and the grid constructed at the candidate's location is used.

\subsection{Optimisation procedure}
\label{sect:maximum}

We have implemented three algorithms to perform a fine search for the maximum of the $\F$-statistic in the \texttt{Followup} procedure described above: (i)  {\it Simplex} method \citep{Spendley1962, Nelder1965} - non-derivative optimisation method based on the simplex (triangle in arbitrary number of dimensions), which relocates and adapts to the local features of the parameter space, and contracts itself toward extremum direction. (ii) {\it Mesh Adaptive Direct Search} (MADS), introduced by \citet{Audet2006} - a grid-based algorithm, which explores parameter space, changing its grid-mesh size, depending on the distance from the extremum. Application of this algorithm to the CGW searches can be found in \citet{Shaltev2013}. (iii) {\it Inverted Mesh Adaptive Direct Search} (invMADS) algorithm - our modification of algorithm (ii) which makes optimisation procedure more suitable for the four-dimensional $\F$-statistic case.

\begin{figure}
	\centering
	\includegraphics[width=\textwidth]{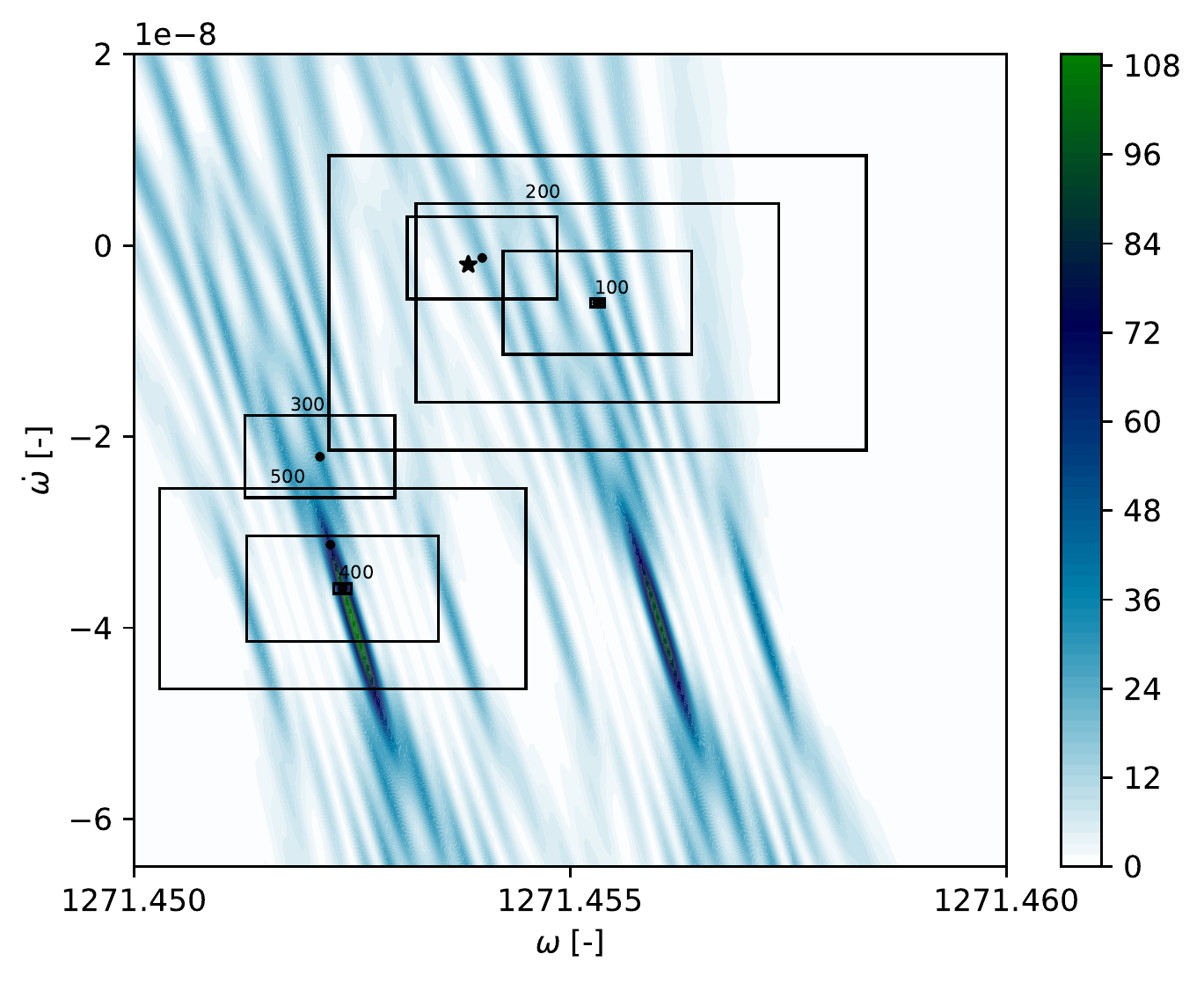}
    \caption{A visualisation of the invMADS algorithm applied to the $\F$-statistic maximum search. Frequency-spindown cut through the four-dimensional space is shown. Initial seed is denoted with the star. Subsequent seeds are denoted with dots and  meshes around them by rectangles. The numbers are the step numbers of the procedure. Contour plot of the $\F$-statistic for the injected noise-free signal is shown in the background. Colour bar encodes values of the $\F$-statistic.}
    \label{fig:mads}
\end{figure}

A visualisation of the simplex algorithm can be found in on Figure 1 of \citet{krolak1999}.
Schematic visualisation of the invMADS algorithm is shown on Fig.~\ref{fig:mads}. The procedure starts from the point with the highest value of $\F$-statistic evaluated on the optimal grid. This {\it initial seed} point is denoted as a star marker on the plot.
Around this point a four-dimensional (two-dimensional on the plot) hypercube is constructed, and the $\F$-statistic is evaluated on the edges, vertices, faces and cells of the hypercube. If the initial seed has a higher value of the $\F$-statistic than on the hypercube points, the mesh is expanded around the initial seed, and the evaluation procedure is repeated. If algorithm finds a higher value of the $\F$-statistic, the seed is changed to that point and the procedure is repeated, starting from the initial size of the hypercube. 
 
Original MADS method as introduced by \citet{Audet2006} is based on dividing the parameter space into relatively large parts and on gradual shrinking toward the direction of the extremum. For the $\F$-statistic where by our follow-up procedure the initial seed point is relatively close to the extremum we have modified the MADS algorithm inverting its idea. The new procedure starts from a small mesh which progressively expands. The pseudo-code of the inverted MADS procedure is as follows:
\newpage
\begin{algorithm}
\caption{Pseudo-code of the inverted MADS procedure}
\KwData{$seed  \longleftarrow$ Initial seed from the optimal grid;\\
$Initial Mesh Size \longleftarrow$ Initial size of the hypercube; \\
$Maximal Mesh Size \longleftarrow$ Maximal (final) size of the hypercube;\\
$Increase \longleftarrow$ How fast hypercube will increase;\\
}
\KwResult{Maximum of the $\F$-statistic}
\While {$Mesh Size < Maximal Mesh Size$ } {
$\F_{seed} \longleftarrow$ Evaluate $\F$-statistic for the seed;\\
$seed + Mesh Size \longleftarrow$ Construct hypercube (mesh) around seed;\\
$\F_{i} \longleftarrow$ Evaluate $\F$-statistic on the vertices, edges, faces and cells of the hypercube;\\
$\F_{max}=max(\F_{i})$\\
\eIf{$\F_{max} > \F_{seed}$}{
$\F_{seed}=\F_{max} \longleftarrow$ Change of the seed\\
$Mesh Size = Initial Mesh Size$\\
}{
$\F_{seed} \longleftarrow$ Unchanged seed\\
$Mesh Size = Mesh Size + Increase$
}
}
\end{algorithm}

\begin{figure}
	\centering
	\includegraphics[width=\textwidth]{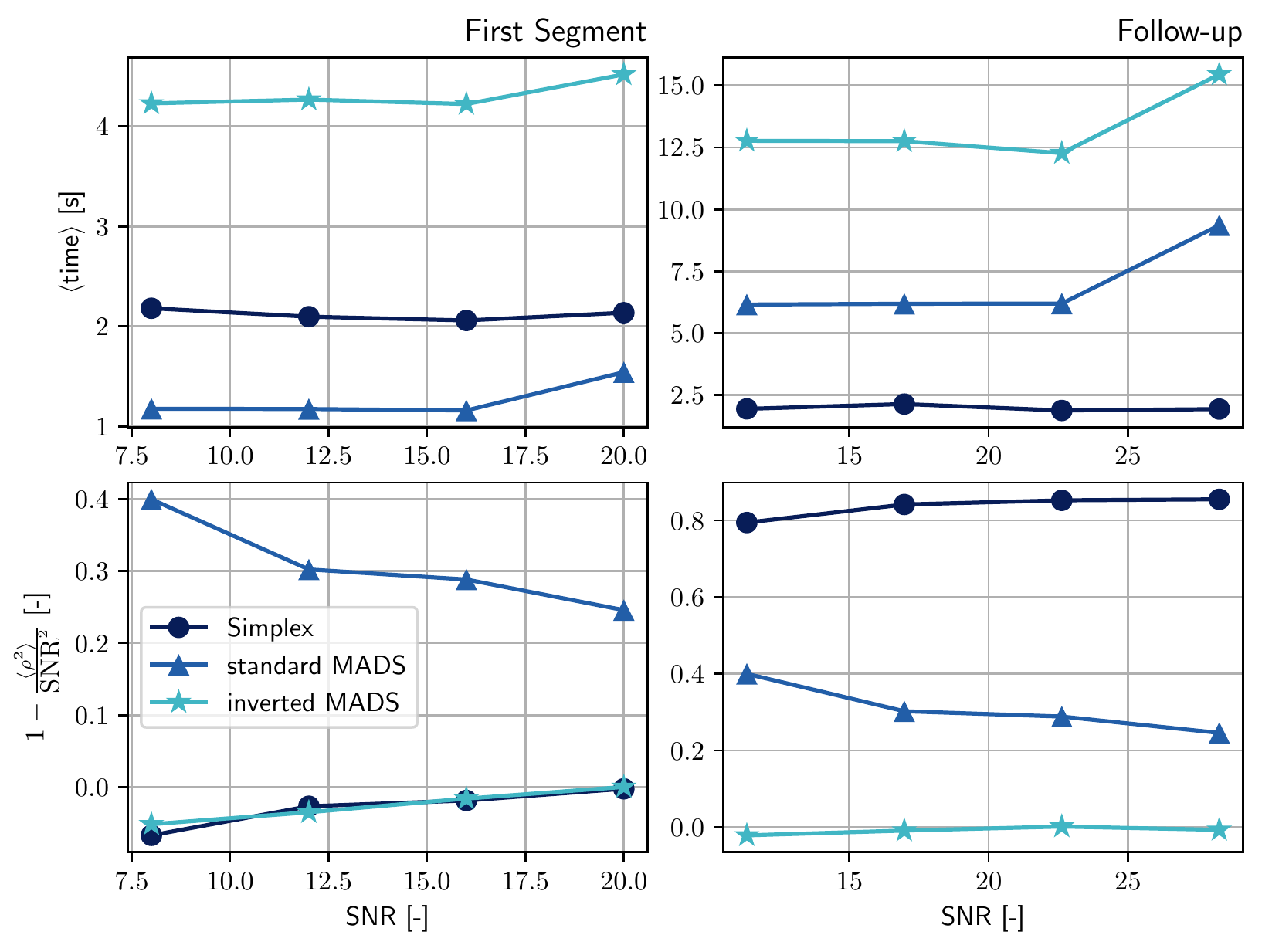}
    \caption{Comparison of the Simplex, MADS, and invMADS algorithms as a function of the signal-to-noise ratio SNR of the injected signal. Left panels show performance for one data segment and right panels for the two data segments joined together. Upper panels show average time (in seconds) of each procedure execution. Lower panels show the quantity 1 - $\langle\rho^2\rangle$/SNR$^2$  which determines the signal-to-noise ratio loss; $\langle\rho^2\rangle$ is the average signal-to-noise the we recover after application of $\F$-statistic maximization procedure. We see that the invMADS procedure is the slowest however it gives the best signal detection especially when we search for maximum in concatenated data segments.}
    \label{fig:mads_vs_simplex}
\end{figure}

Comparison of the three algorithms is shown on Fig.~\ref{fig:mads_vs_simplex}. The only criterion of choosing parameters fixed for each procedure, like for example \textit{Initial Mesh Size} or \textit{Increase} parameters for invMADS (see pseudo-code above), was the good convergence of the results to the theoretical predictions (Cramer-Rao bounds, see Sect.~\ref{sect:test4d} and Fig.~\ref{fig:std}) for a single data segment.
We find essential differences in the algorithms performance for one data segment (left column) and two data segments joined together (right column). For the standard and the inverted MADS the execution time increases several times when switching from the shorter to the longer time series. This is because for the longer time series the $\mathcal{F}$-statistic becomes more narrowly peaked, as was shown on Fig.~\ref{fig:Fstat} - narrower extrema are harder to find in the 4-dimensional space and it is more time-consuming to check whether the extremum is the global one. Execution time for Simplex algorithm is similar for single and concatenated data segments. This can be explained by the bottom panels on Fig.~\ref{fig:mads_vs_simplex}. The algorithm finds correct extrema (SNR loss is close to 0) for a single data segment however for the longer time series, where, as was mentioned above, it is harder to find global extremum, algorithm is as fast as previously but it gets stuck in some local peaks (SNR loss for Simplex in the Follow-up stage is very large ($\sim 0.8$) ). Standard MADS procedure does not fit our purposes: it generates relatively large SNR loss for a single data segment as well as in the Follow-up stage. The most promising algorithm in our test is inverted MADS. SNR loss for our inverted MADS both for shorter and concatenated data segments remains stable and is close to 0.
Disadvantage of that method is that it is slow. However this is not a problem as the accuracy and not the computational speed is the main requirement for the follow-up procedure. Additionally, invMADS can be easily sped-up, e.g. by adding parallelisation to the procedure 
whereas Simplex is very hard to parallelise.

\section{Tests in Gaussian noise}
\label{sect:testsGN} 

In this chapter we present implementation of the \texttt{Followup} procedure in computer codes 
and we carry out Monte Carlo simulations to present efficiency of our follow-up method.
We present two simulations for the case of CWG signals buried in white Gaussian noise. First simulation is for a signal that depends only on 4 parameters: amplitude $h_o$, phase $\phi_o$, angular frequency $\omega$, and frequency derivative $\dot{\omega}$. The $\F$-statistic for such a signal depends only on two parameters - $\omega$ and $\dot{\omega}$. The second simulation is for the general signal presented in Section \ref{sect:polgraw} that depends on eight parameters. The main principle of our follow-up procedure is to coherently join the data segments into a longer time series to increase detection sensitivity.

\begin{figure}
	\centering
	\includegraphics[width=\textwidth]{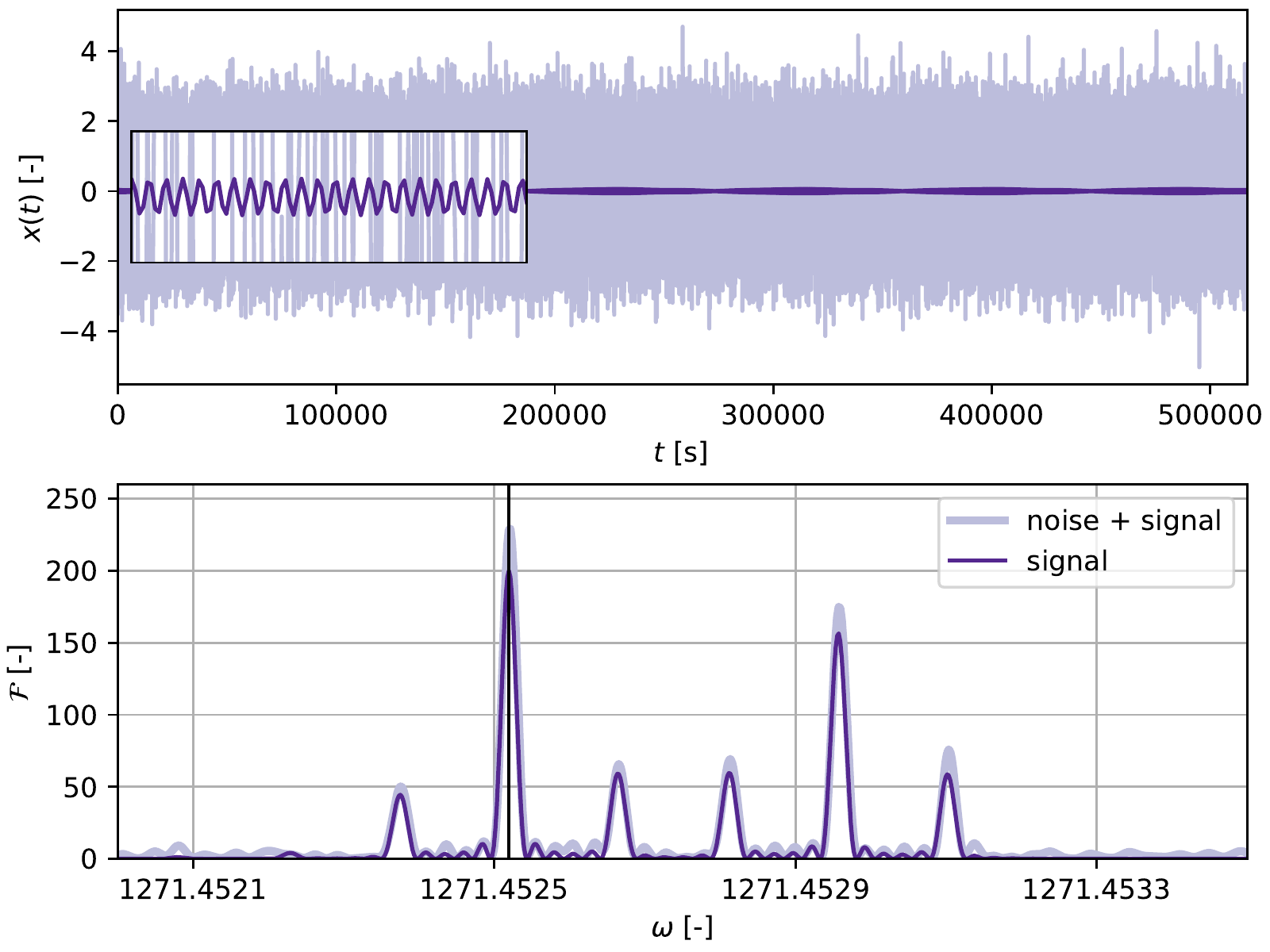}
    \caption{Upper panel: artificial signal (SNR$=20$, dark purple) added to the white Gaussian noise of mean 0 and variance 1 (light purple). Length of the time series (x axis) is 6 days long. Lower panel: the $\F$-statistic for the data presented on the upper panel. The function exhibits subsidiary maxima that are due to amplitude modulations. Dark line corresponds to the $\F$-statistic shape for the pure signal without the noise, while lighter line unveils effect of existence of the noise: heights, shapes and positions of the extrema are changed. Frequency of the injected signal is denoted with the vertical line.  }
    \label{fig:noise_and_signal}
\end{figure}

In Fig.~\ref{fig:noise_and_signal} we plot sample data
and the $\F$-statistic used in our follow-up procedure.
The data consists of the signal given by Eq. \ref{sig} 
with the optimal signal-to-noise ratio $\rho=20$ (Eq.\ref{snr})   added to white Gaussian noise (upper panel). We see the signal
is deeply buried in the noise. The lower panel shows the 
$\F$-statistic (Eq.\ref{eq:Fstat}). We see that $\F$-statistic has several subsidiary maxima. The maxima are due to the amplitude modulation and are separated by multiples of sidereal day frequency. The amplitude modulation functions have 2 harmonics of sidereal frequency. The $\F$-statistic involves modulus squared of
the amplitude demodulated data and thus it has in general 4 harmonics on both sides of the main peak. The relative amplitudes of the subsidiary maxima depend on the declination of the source and for the declination of the signal used in Fig.~\ref{fig:noise_and_signal} only 5 subsidiary maxima are visible. 

\begin{figure}
	\centering
\includegraphics[width=\textwidth]{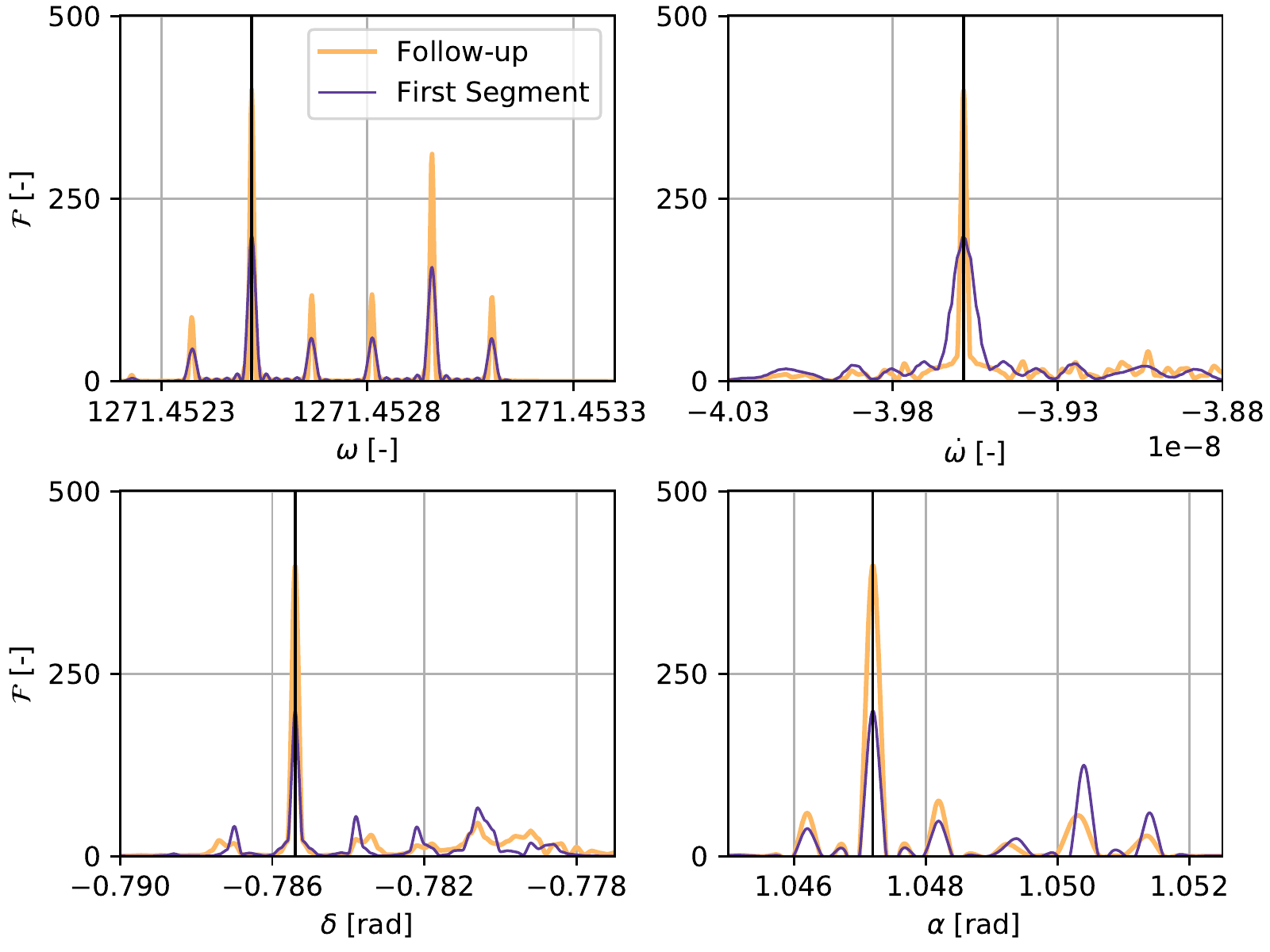}
    \caption{$\F$-statistic for the single (purple) and concatenated (orange) data segments. Each subplot shows one-dimensional cut in the place of added artificial signal with SNR$= 20$, through four-dimensional parameter space. Parameters of the injected signal are denoted with the black vertical line.}
    \label{fig:Fstat}
\end{figure}

In Fig.~\ref{fig:Fstat} we show the $\F$-statistic as
a function of frequency for one segment and two segments joined together. We see that for two segments joined together
the amplitude of the $\F$-statistic increases and width of the
function decreases; the $\F$-statistic becomes more narrowly peaked.

\subsection{Two-dimensional case}
\label{sect:test2d}

In this case the signal $s(t)$ is given by
\be
\label{eq:sig2}
s(t) = h_o \cos(\omega t + \dot{\omega} t^2 + \phi_o)
\ee
We assume that the signal present in the data $x(t)$ is additive i.e.
\be
x(t) = n(t) + s(t),
\ee
where $n(t)$ is Gaussian white noise of mean $0$ and variance $\sigma^2$. The $\F$-statistic in this case is given by
\be
\label{eq:Fstat2}
\F =
\frac{|\sum^N_{t=1} x(t)\, \exp[i(\omega t + \dot{\omega} t^2)]|^2}{N \sigma^2},
\ee
where $x_k(t), t = 1, ... , N$ are $N$ data points.
 
For signal (\ref{eq:sig2}), the signal-to-noise ratio $\rho$ is approximately given by
\be
\rho = \frac{h_o}{\sigma} \sqrt{\frac{N}{2}} 
\ee
and the Fisher matrix projected on the two-dimensional space spanned by angular frequency
and the spin down parameter is given by
\be
\label{eq:Mmat2}
\Gamma = \rho^2
\left(\begin{array}{cc}
  \frac{1}{12} N^2  &  \frac{1}{12} N^3 \\[1ex]
  \frac{1}{12} N^3  &  \frac{4}{45} N^4  
      \end{array}\right).
\ee
Consequently the variances of the estimators of
$\omega$ and $\dot{\omega}$ approximately read
\bea
\label{eq:sigom}
\sigma_{\omega} &=& \frac{8\sqrt{3}}{\rho N}  \\
\label{eq:sigom1}
\sigma_{\dot{\omega}} &=& \frac{6\sqrt{5}}{\rho N^2}
\eea 

Next we present construction of the optimal grid outlined in Section \ref{sect:OptGrid} for the case of the signal given by Eq.~\ref{eq:sig2}. Here the intrinsic parameter space is two-dimensional and spanned by angular frequency $\omega$ and the spin down $\dot{\omega}$. 
The thinnest covering in the two-dimensional space is the hexagonal lattice denoted by $A_2^*$ \citep{Con}. 

The generator matrix of this lattice is given by
\begin{equation}
  \label{eq:A2gen}
  M_o = R\sqrt{3}
  \left(\begin{array}{cc}  
      1&0\\
      \frac{1}{2}&\frac{\sqrt{3}}{2}
          \end{array}\right),
\end{equation}
where $R$ is the covering radius of the lattice.
The \emph{reduced} Fisher matrix $\tilde{\Gamma}$
for signal (Eq.~\ref{eq:sig2}) reads
\begin{equation}
  \label{eq:gamma2d}
  \tilde{\Gamma}=  \frac{\Gamma}{\rho^2}.
\end{equation}
The generator matrix $M_2$ transformed to the parameters $\omega_0$ and $\omega_1$ is obtained from the equation
\be
M_2 = M_o \cdot V'\cdot \sqrt{E},
\ee
where $V$ is the matrix whose columns are eigenvectors of the reduced Fisher matrix $\tilde{\Gamma}$, $V'$ denotes transpose of matrix $V$  and $E$ is the diagonal matrix whose diagonal components are eigenvalues of $\tilde{\Gamma}$.
To characterize the density of the grid we introduce parameter $MM$ defined as
\be
MM = \sqrt{1 - R^2}.
\ee
The denser the grid the closer the value of the parameter $MM$ to $1$.  

For our simulations we generate a set of data containing
$2N$ data points. The data consists of Gaussian noise of mean 0 and variance 1 and signal (Eq.~\ref{eq:sig2}) added to it.
We first divide the data into two segments of $N$ data points and  perform the search and estimation of the parameters in the two segments for various signal-to-noise ratios according to the first step of the \texttt{Followup} procedure
presented in Sect. \ref{sect:steps}. 
To obtain the signal with a specified SNR we scale the amplitude $h_o$ appropriately. 
We first verify whether the detectability of the signal agrees with the theoretical one. In the case of signal present $2\times\F$-statistic defined by Eq. \ref{eq:Fstat2} has a non-central $\chi^2$ distribution of 2 degrees of freedom with non-centrality parameter equal to $\rho^2$.
Consequently the theoretical mean ($\mu$) and variance ($\Sigma^2$)  for $2\times\F$ are given by Eqs.\ref{eq:mu} with $n = 2$.

\begin{figure}
	\centering
	\includegraphics[width=\textwidth]{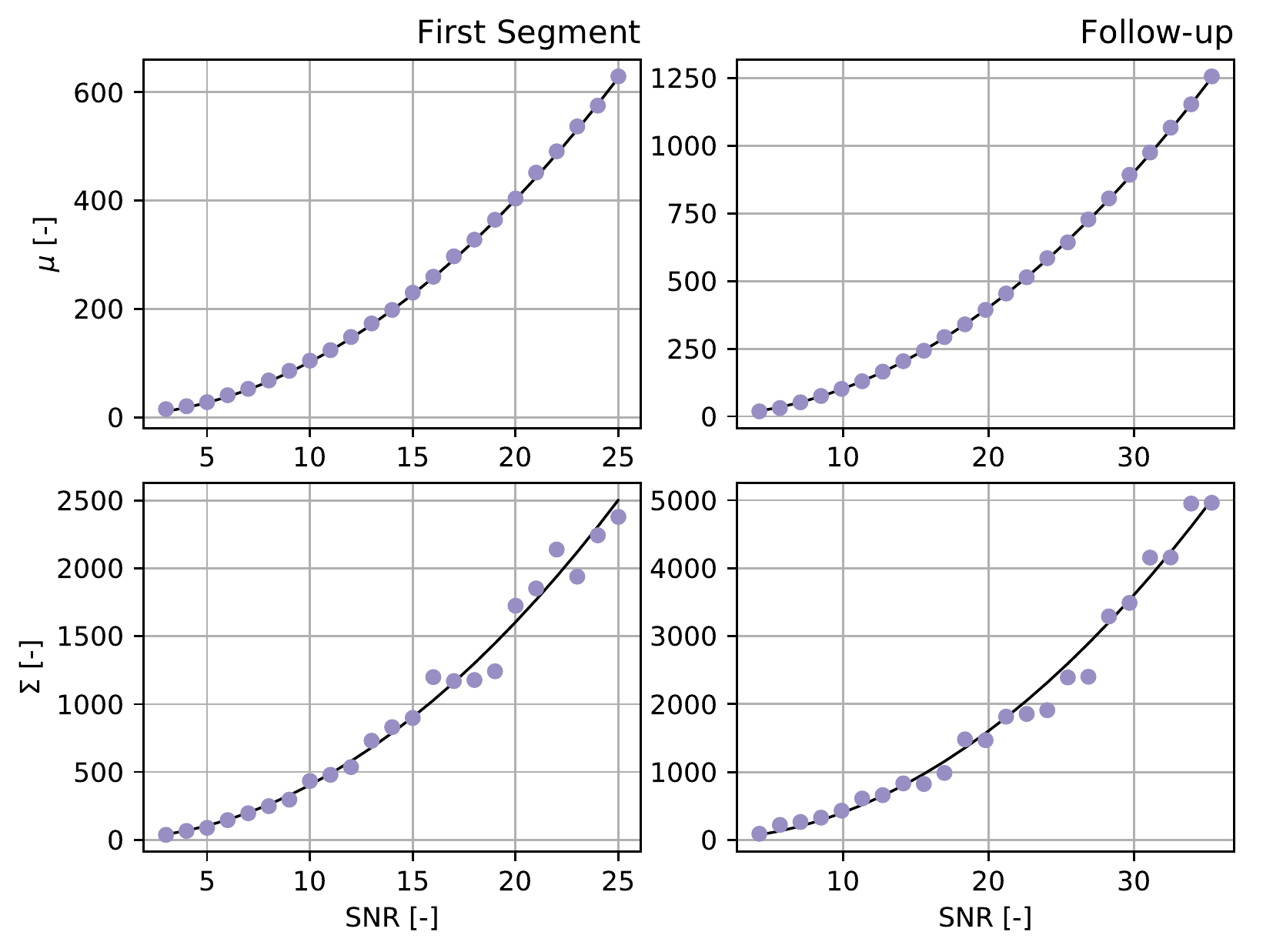}
    \caption{Top two panels show the mean of the $\F$-statistic obtained from the simulations (dots) in comparison to the theoretical one (continuous line). The left panel is 
for 1st data segment of $N$ data points and the right panel is for two segments joined together after the follow-up procedure. The bottom two panels show the results for the variance of the $\F$-statistic.}
    \label{fig:detect}
\end{figure}

In Figure \ref{fig:detect} we plot the mean and variance
of $\F$-statistic obtained from our simulations as functions of the SNR and compare them with the theoretical ones. 
For each SNR we perform 250 simulations.
The good agreement for the mean and variance after the \texttt{Followup} means  that the \texttt{Followup} procedure achieves the theoretical increase  in the signal-to-noise ratio and leads to best possible increase in signal detectability.

\begin{figure}
	\centering
	\includegraphics[width=\textwidth]{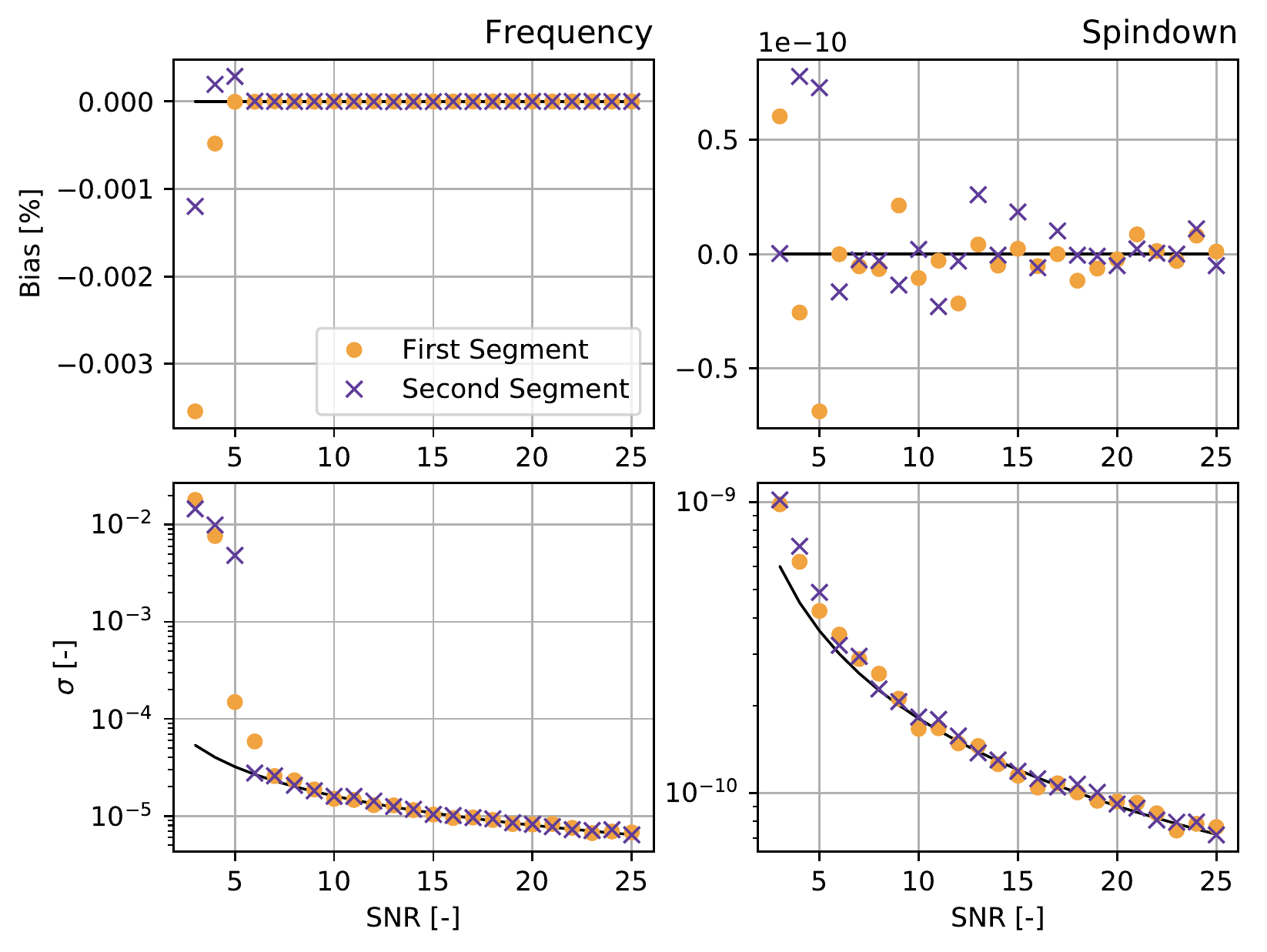}
    \caption{Top two panels show bias of the estimators of frequency (left panel) and spin down (right panel) obtained form the simulations for the two segments of $N$ data points analysed. The bottom two panel show standard deviations 
obtained from the simulation in comparison with predictions of the Fisher matrix.}
    \label{fig:2d1seg}
\end{figure}

\begin{figure}
	\centering
	\includegraphics[width=\textwidth]{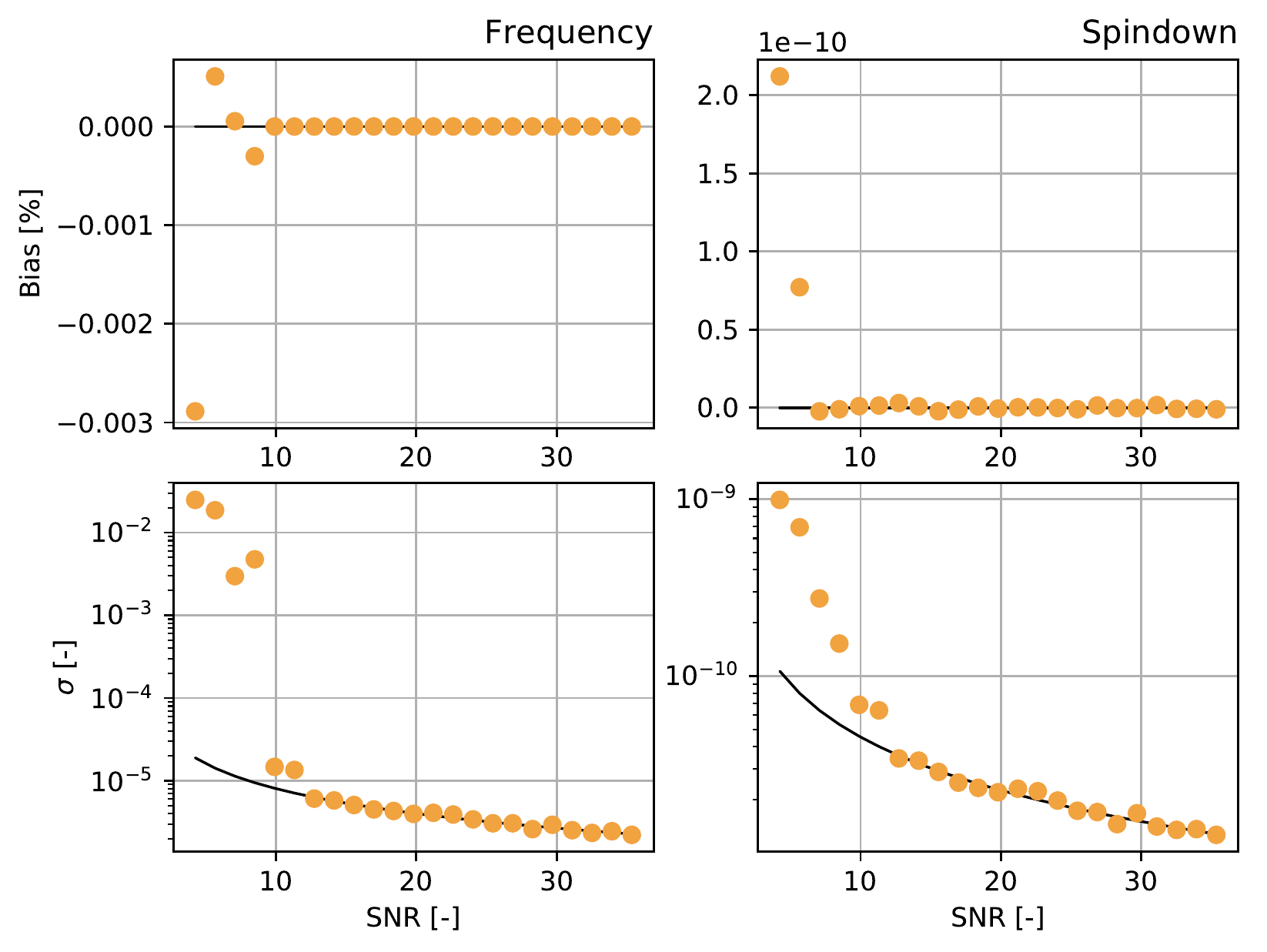}
    \caption{The results of the simulation are presented as in Fig.~\ref{fig:2d1seg} when two segments are coherently joined together and the follow-up procedure is performed.}
    \label{fig:2dfollowup}
\end{figure}

The results of the estimation of parameters are presented in Fig. \ref{fig:2d1seg}.  We calculate means and standard deviations for 250 estimators of parameters $\omega$ and \textbf{$\dot{\omega}$} that we obtain and we compare them with the true values of these parameters and the standard
deviations predicted by the Fisher matrix and given by
Eqs.~\ref{eq:sigom} and \ref{eq:sigom1}. 
The results of the comparison are presented on Fig.~\ref{fig:2d1seg} and Fig.~\ref{fig:2dfollowup}. 
The bias for a given parameter $\theta$ expressed in \% is defined as
\begin{equation}
\textrm{Bias [\%]} = 
100 \cdot \left( 1 - \frac{\bar{\theta}}{\theta} \right),
\label{eq:bias_percentage}
\end{equation}
where $\bar{\theta}$ is the mean value of the estimators of the parameter $\theta$ and $\theta$ is its true value.

We note a very good agreement of results our simulations with the Fisher matrix above $\rho = 7$. We then carry out the second stage of the \texttt{Followup} procedure i.e. the search for the signal in the whole $2 N$ data set and we again see a good agreement with the Fisher matrix predictions above  $\rho = 12$ (see Fig. \ref{fig:2dfollowup}). In both stages of the procedure we used the optimal grid constructed above with parameter $MM = 0.98$.

\subsection{General four-dimensional case}
\label{sect:test4d}

\begin{figure}
	\centering
	\includegraphics[width=\textwidth]{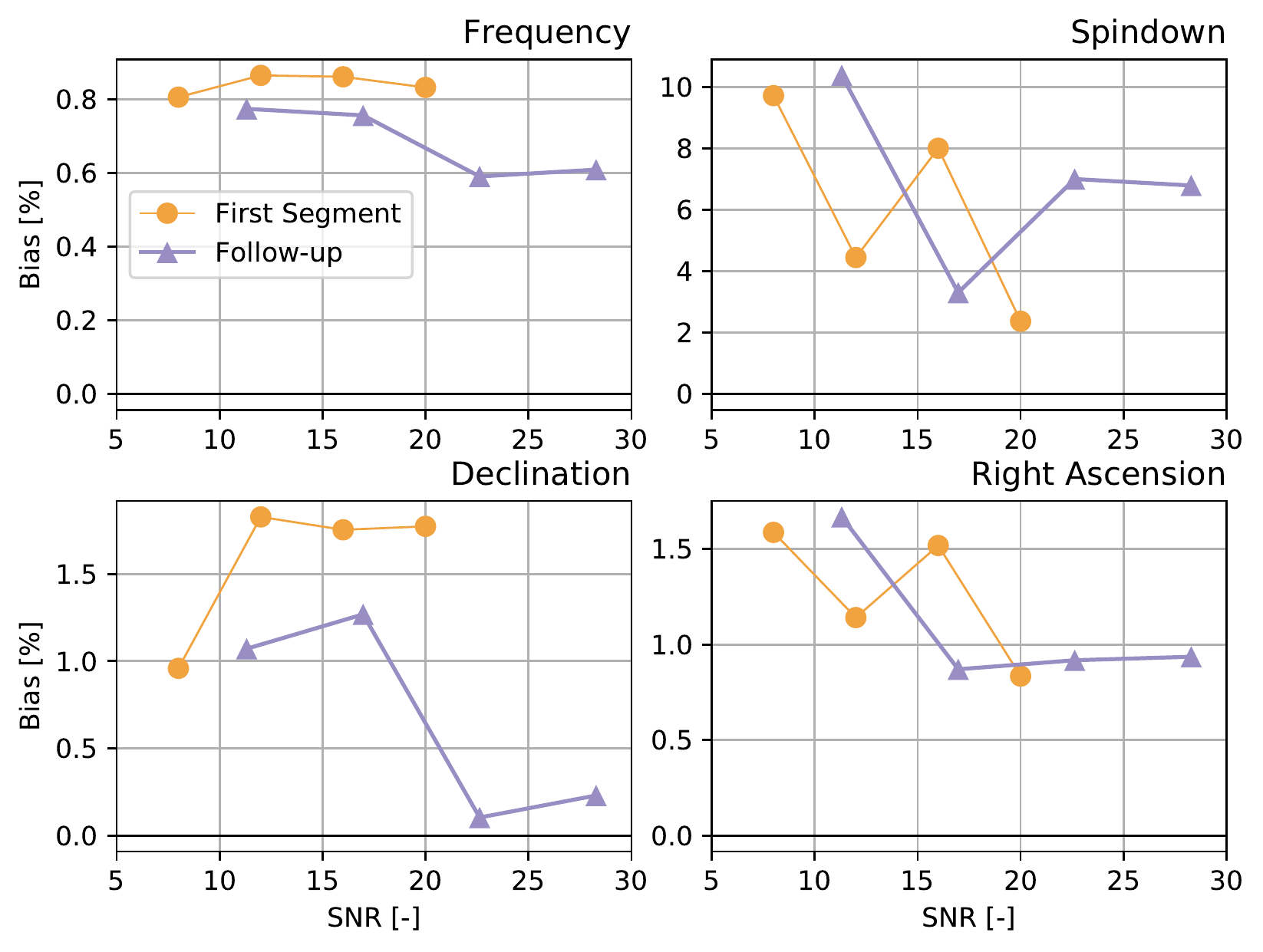}
    \caption{Bias of the estimators of four parameters of the signal for one data segment and two data segments joined together versus the signal-to-noise ratio of the injected signal.}
    \label{fig:bias}
\end{figure}

\begin{figure}
	\centering
	\includegraphics[width=\textwidth]{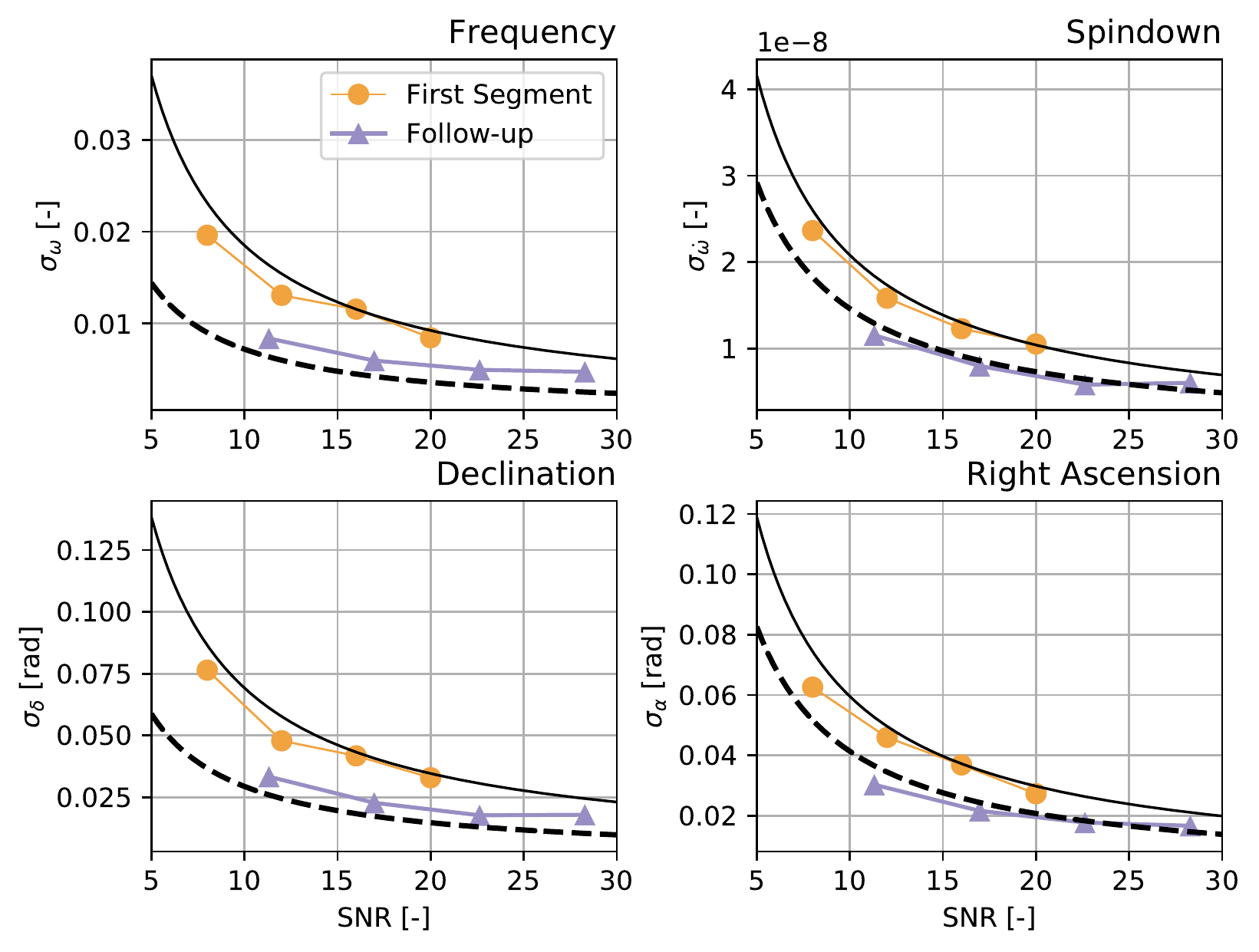}
    \caption{Standard deviations of the parameter estimators
for one and two segments joined together. The continuous and dashed line are predictions from the Fisher matrix for one segment and two segments respectively.}
    \label{fig:std}
\end{figure}

Our code allows us to inject artificial signal with an arbitrary amplitude or equivalently with an arbitrary SNR. 
To check performance of our follow-up code we injected signals with different values of SNR ranging from 8 to 20. Signal was buried in white Gaussian noise.
To make our simulation more realistic and applicable to real search where we analyse many consecutive time segments
we generated data consisting of six data segments. 
In each data segment we search for the signal using our two-step procedure: first a grid search using the optimal grid constructed in Sect.~\ref{sect:OptGrid} and then a fine search using the Simplex algorithm. We than calculate the mean values of the parameters of the signal obtained from the fine search weighted by the estimated SNRs of the signal in each frame.
These mean values constitute the initial values of the parameters for the follow-up procedure. We then apply the follow-up procedure to the first two data segments that we join together using the invMADS algorithm 
(Sect.~\ref{sect:maximum}) with the initial values of the parameters. We repeat whole procedures 250 times, for different realisations of the Gaussian noise. 
The results of the simulation are show in Fig.~\ref{fig:bias} and Fig.~\ref{fig:std}. In Fig.~\ref{fig:bias} we show the bias of the estimators of the four parameters and in 
Fig.~\ref{fig:std} the standard deviation in comparison with the predictions of the Fisher matrix. For the follow-up case we have plotted the bias and the standard deviation as a function of signal-to-noise ratio
for the two segments joined together which is a factor of
$\sqrt{2}$ larger than for one segment. We note a satisfactory performance of the follow-up procedure. Namely the bias of the estimators decreases as we move from one segment to two segments joined together and standard deviations of the estimators improve in accordance with the Fisher matrix predictions.

\begin{figure}
	\centering
	\includegraphics[width=\textwidth]{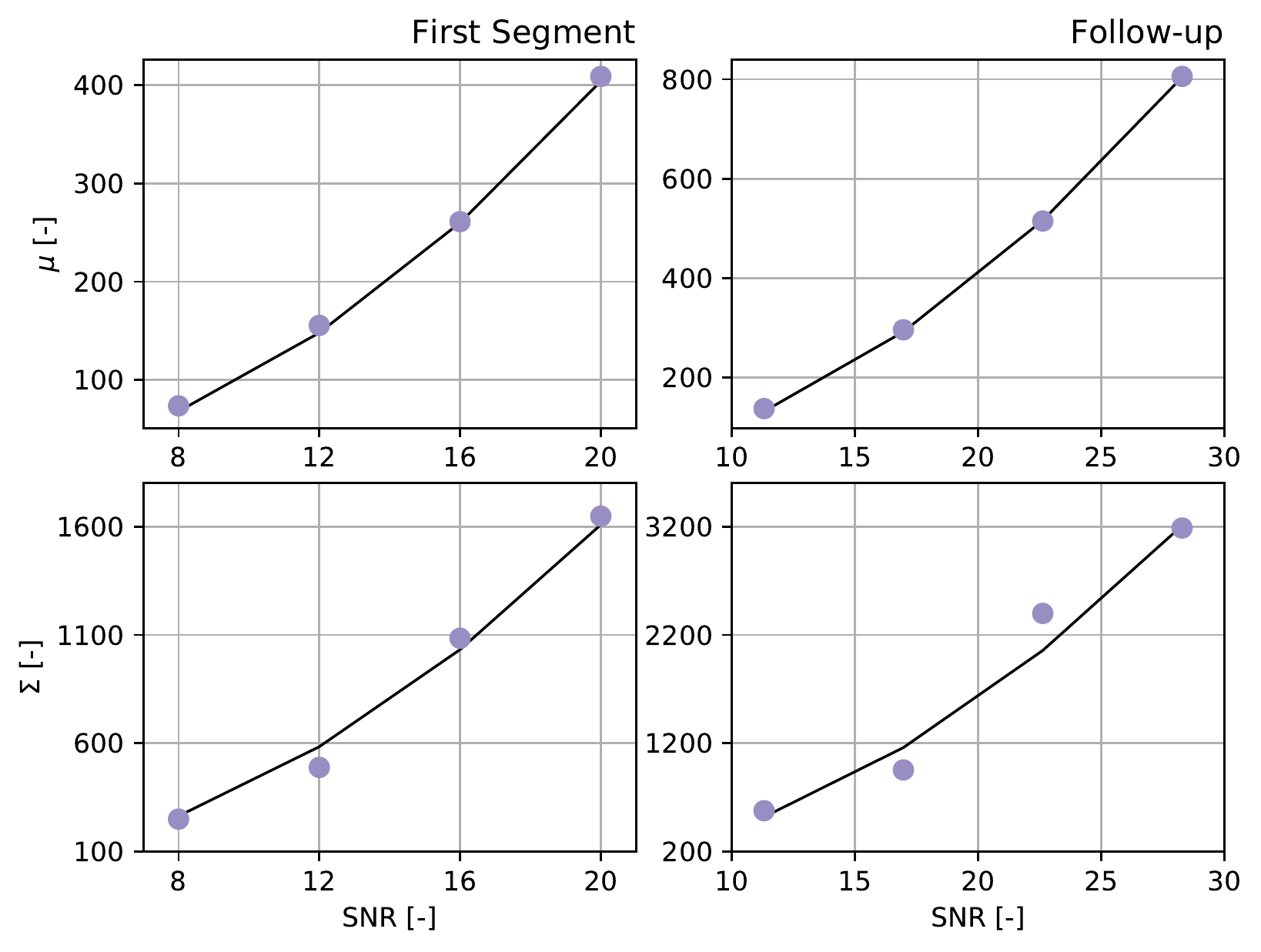}
    \caption{Comparison of means (upper panels) and variances (lower panels) of the $\F$-statistic obtained from the simulations in Gaussian noise (dots) with theoretical expectations (solid lines). The left panel is for one data segment and the right panel is for two segments joined together.}
    \label{fig:det}
\end{figure}

Similarly, like in the two-dimensional case (Sect.~\ref{sect:test2d}), we also verify whether the detectability of the signal agrees with the theoretical one. In the 4-dimensional case $2\times\F$-statistic has a non-central $\chi^2$ distribution of 4 degrees of freedom with the non-centrality parameter equal to $\rho^2$. Consequently the theoretical mean ($\mu$) and variance ($\Sigma^2$)  for $2\times\F$ are given by Eq.~\ref{eq:mu} with $n = 4$.
In Fig.~\ref{fig:det} we present the comparison of our simulations with theoretical expectations for an array of signal-to-noise ratios.
 
\begin{figure}
	\centering
	\includegraphics[width=\textwidth]{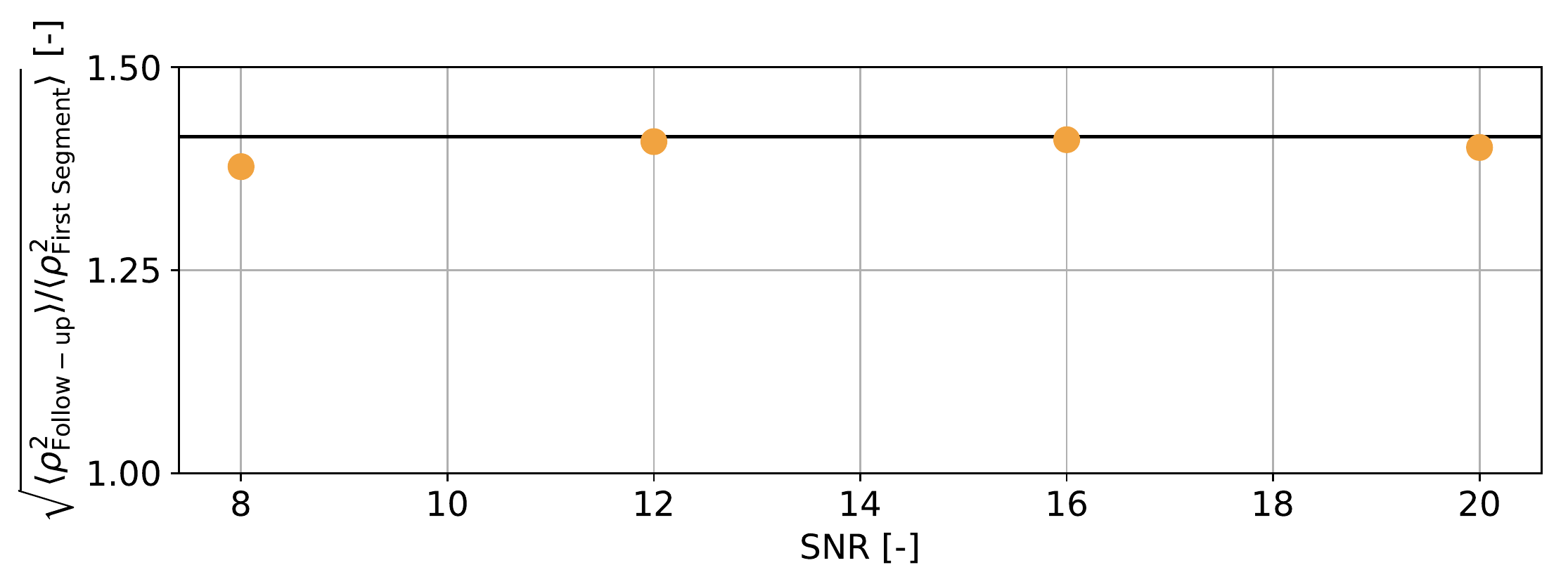}
    \caption{Mean increase of signal-to-noise ratio $\rho$ with observational time. Theoretically, concatenation of the two data segments into one long data segment should result as improvement of $\rho$ by factor of $\sqrt{2}$ (horizontal line). Results of our tests are denoted with dots.}
    \label{fig:det_obs}
\end{figure}

We have also investigated how on average the SNR
of the injected signal increases in the course of our follow-up procedure. The increase is measured by the quantity $\sqrt{\frac{\langle\rho_2^2\rangle}{\langle\rho_1^2\rangle}}$ where $\langle\rho_1^2\rangle$ and $\langle\rho_2^2\rangle$ are the averaged squared SNRs of the signal in two segments joined together and one segment respectively. The obtained results are shown in Fig.~\ref{fig:det_obs}. We see
a good agreement with the expected increase of SNR by $\sqrt{2}$.

\section{Conclusions}
\label{sect:summ}

In this paper we have presented a follow-up procedure 
for the analysis of the candidate signals obtained from 
the \texttt{Time-Domain $\F$-statistic} pipeline. The aim
of the procedure is to verify whether a given candidate signal can be of astrophysical signal. The procedure is based on the assumption that a true signal is coherent and always present in the data and thus signal-to-noise ratio of the signal should increase as square root of the observation time.

The basic tool of the procedure is the $\F$-statistic. We have described data analysis methods and algorithms used in the procedure.  They involve construction of optimal grids of templates over which the $\F$-statistic is evaluated and optimization algorithms to find accurately the global maximum
of the $\F$-statistic.  We have presented detailed steps of
the procedure. The follow-up would proceed hierarchically -
we first join two data segments and analyse them coherently,
then concatenate data segments into four data segments and so on. In practice the candidate events from our pipeline are
results of search for coincidences in  $L$ data segments.
A significant candidate need not be coincident in all $L$ data segments. Nevertheless by our basic assumption a candidate of astrophysical origin must be always present in the data. Thus in our follow-up we proceed with the steps outlined in Section \ref{sect:steps} to search for signal in two consecutive segments as if the signal was present in both segments.

We have performed Monte Carlo simulations of the  \texttt{Followup} by injecting artificial signal into white Gaussian noise. We have tested the procedure  in two cases.
Firstly, for a simple two-dimensional model when only two parameters - frequency $\omega$ and  spindown $\dot{\omega}$
are unknown. Then we have performed tests for a general four-dimensional signal that is modulated by the motion of the detector and depends also on the position of the source in the sky. Our test involves comparison of the estimated parameters of the signal with predictions of the Fisher matrix. We find a satisfactory agreement. Thus we expect that the procedure will become an integral part of the \texttt{Time-Domain $\F$-statistic} pipeline and will be applied to the analysis of real data. As the methods presented here are quite general, they can be applied to any other continuous GW signal, e.g. in the searches for QCD axion clouds bound to black holes \citep{Arvanitaki2015}. 

\section*{Acknowledgements}
The work was supported in part by the Polish National Science Centre grants no. 2016/22/E/ST9/00037, 2017/26/M/ST9/00978 and 2018/28/T/ST9/00458.

\end{document}